# Web Content Classification: A Survey

Prabhjot Kaur
*Dept. of Computer Science and Engg,
Sri Guru Granth Sahib World University,
Fatehgarh Sahib, Punjab, India*

**ABSTARCT:** *As the information contained within the web is increasing day by day, organizing this information could be a necessary requirement.The data mining process is to extract information from a data set and transform it into an understandable structure for further use. Classification of web page content is essential to many tasks in web information retrieval such as maintaining web directories and focused crawling.The uncontrolled type of nature of web content presents additional challenges to web page classification as compared to the traditional text classification ,but the interconnected nature of hypertext also provides features that can assist the process. In this paper the web classification is discussed in detail and its importance in field of data mining is explored.*

**Keywords-** *Data mining, Web page Classification, Feature Selection, Classification.*

## I. INTRODUCTION

Data mining or "Knowledge Discovery and Data Mining" process an interdisciplinary subfield of computer science is the computational process of discovering patterns in large data sets involving methods at the intersection of statistics and artificial intelligence etc.The goal of the data mining process is to extract information from a data set and transform it into an understandable structure for further use. Apart from the raw analysis step it involves database and data management aspects data pre-processing model and inference considerations, interestingness metrics, visualization and online updating[12].

The actual data mining task is the automatic or semi-automatic analysis of large quantities of data to extract previously unknown interesting patterns such as groups of data records (cluster analysis) unusual records (anomaly detection) and dependencies (association rule mining). This involves using database techniques such as spatial indices. The patterns can then be seen as a kind of summary of the input data and may be used in further analysis for example in machine learning and predictive analytics. Like the data mining step might identify multiple groups in the data which can then be used to obtain more accurate prediction results by a decision support system.The data collection, data preparation and even not the result interpretation and reporting are part of the data mining step but do belong to the overall KDD process as additional steps.

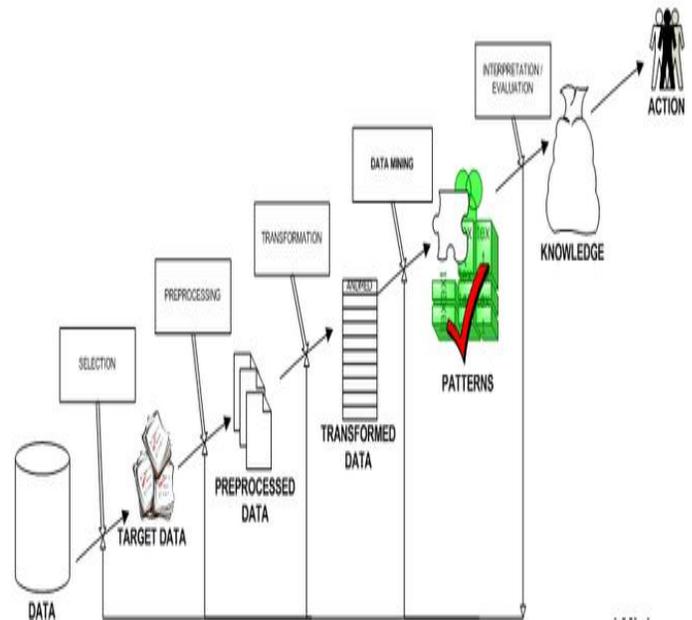

Fig.1 Steps in Data Mining[12].

### A. Data mining applications

Data mining is increasingly popular because of its substantial contribution. It can be used to control costs as well as contribute to revenue increases. Many organizations are using data mining to help manage all phases of the customer life cycle including acquiring new customers increasing revenue from existing customers and retaining good customers. Determining characteristics of good customers (profiling) a company can target prospects with similar characteristics. After profiling customers who have bought a particular product it can focus attention on similar customers who have not bought that product.By profiling customers who have left a company can act to retain customers who are at risk for leaving  because it is usually far less expensive to retain a customer than acquire a new one.

Data mining offers value across a broad spectrum of industries.The telecommunications and credit card companies are two of the leaders in applying data mining to detect fraudulent use of their services. The insurance companies and stock exchanges are also interested in applying this technology to reduce fraud. The medical applications are another fruitful area where data mining can be used to predict the effectiveness of surgical procedure or medications. The companies active in the financial markets use data mining to determine market and industry characteristics as well as to predict individual company and stock performance. The retailers are making more use of data mining to decide which products to stock in particular stores, as well as to assess the effectiveness of promotions and coupons. The pharmaceutical firms are mining large databases of chemical compounds and of genetic material to discover substances that might be candidates for development as agents for the treatments of disease.

B.  Six common classes of tasks in Data Mining:

•       Anomaly detection (Outlier/change/deviation detection) – The identification of unusual data records that might be interesting or data errors that require further investigation.

•       Association rule learning – Searches for relationships between variables. Like in a supermarket might gather data on customer purchasing habits. By using association rule learning the supermarket can determine which products are frequently bought together and use this information for marketing. This is sometimes referred to as market basket analysis.

•        Clustering – It is the task of discovering groups and structures in the data that are in some way or another "similar" without using known structures in the data.

•        Classification – It is the task of generalizing known structure to apply to new data. Like an e-mail program might attempt to classify an e-mail as "legitimate" or as "spam".

•        Regression – Attempts to find a function which models the data with the least error.

•       Summarization – providing a more compact representation of the data set, including visualization and report generation[12].

## II.  WEB PAGE CLASSIFICATION

With the rapid growth of the World Wide Web (WWW) there is an increasing need to provide automated assistance to Web users for Web page classification and categorization. Such assistance is helpful in organizing the vast amount of information returned by keyword-based search engines or in constructing catalogues that organize web documents into hierarchical collections. However it is difficult to meet without automated web-page classification techniques due to the labor-intensive nature of human editing. On a first glance web-page classification can borrow directly from the machine learning literature for text classification. On closer examination however the solution is far from being so straightforward[8]. The web pages have their own underlying embedded structure in the HTML language. They typically contain noisy content such as advertisement banner and navigation bar. If a pure-text classification method is directly applied to these pages, it will incur much bias for the classification algorithm making it possible to lose focus on the main topics and important content. Thus a critical issue is to design an intelligent preprocessing technique to extract the main topic of a WebPage[10].

The general problem of web page classification can be divided into multiple sub-problems such as functional classification and other types of classification. The subject classification is concerned about the subject or topic of a web page. Based on the number of classes in the problem classification can be divided into binary classification and multi-class classification where binary classification categorizes instances into exactly one of two classes multi-class classification deals with more than two classes. Depending upon the number of classes that can be assigned to an instance classification can be divided into single-label classification and multi-label classification. What happens in single-label classification one and only one class label is to be assigned to each instance while in multi-label classification more than one class can be assigned to an instance. Like if a problem is multi-class say four-class classification it means four classes are involved say Arts, Business, Computers and Sports. This can be either single-label where exactly one class label can be assigned to an instance or multi-label where an instance can belong to any one two or all of the classes. Based on the type of class assignment classification can be divided into hard classification and soft classification.In the system of Hard classification an instance can either be or not be in a particular class without an intermediate state on the other soft classification an instance can be predicted to be in some class with some likelihood often a probability distribution across all classes.

Based on the organization of categories the web page classification can also be divided into flat classification and hierarchical classification.Flat classification categories are considered parallel i.e. one category does not supersede another.On the other hand in hierarchical classification the categories are organized in a hierarchical tree-like structure in which each category may have a number of subcategories.

Classification of web content is different in some aspects as compared with text classification. The uncontrolled nature of web content presents additional challenges to web page classification as compared to traditional text classification. The web content is semi structured and contains formatting information in form of HTML tags. A web page consists of hyperlinks to point to other pages. This interconnected nature of web pages provides features that can be of greater help in classification. First all HTML tags are removed from the web pages including punctuation marks. The next step is to remove stop words as they are common to all documents and does not contribute much in searching. In most cases a stemming algorithm is applied to reduce words to their basic stem. One such frequently used stemmer is the Porter's stemming algorithm. Machine learning algorithms are then applied on such vectors for the purpose of training the respective classifier. The classification mechanism of the algorithm is used to test an unlabelled sample document against the learnt data. In this approach user deal with home pages of organizational websites. [9]A neatly developed home page of a web site is treated as an entry point for the entire web site. It represents the summary of the rest of the web site. Many URLs link to the second level pages telling more about the nature of the organization. The information contained the title, meta keyword, meta description and in the labels of the A HREF (anchor) tags are very important source of rich features. In order to rank high in search engine result site promoters pump in many relevant keywords. Most of the homepages are designed to fit in a single screen. The factors discussed above contributed to the expression power of the home page to identify the nature of the organization[11].

### III. RELATED WORK

In the paper [1] they have introduced that increase in the amount of information on the Web has caused the need for accurate automated classifiers for Web pages to maintain Web directories and to increase search engine performance. Every tag and every term on each Web page can be considered as a feature there is a need for efficient methods to select best features to reduce feature space of the Web page classification problem. The aim of this paper is to apply a recent optimization technique namely the firefly algorithm (FA) to select best features for Web page classification problem. The firefly algorithm (FA) is a metaheuristic algorithm, inspired by the flashing behavior of fireflies.Using FA to select a subset of features and to evaluate the fitness of the selected features J48 classifier of the Weka data mining tool is employed. Web KB and Conference datasets were used to evaluate the effectiveness of the proposed feature selection system.Observation is that when a subset of features are selected by using FA, WebKB and Conference datasets were classified without loss of accuracy even more time needed to classify new Web pages reduced sharply as the number of features were decreased.

In the paper [2] they have introduced the Web has become one of the most widespread platforms for information change and retrieval. As it becomes easier to publish documents as the number of users and thus publishers, increases and as the number of documents grows searching for information is turning into a cumbersome and time-consuming operation. Due to heterogeneity and unstructured nature of the data available on the WWW Web mining uses various data mining techniques to discover useful knowledge from Web hyperlinks page content and usage log.The main uses of web content mining are to gather categorize, organize and provide the best possible information available on the Web to the user requesting the information. The mining tools are imperative to scanning the many HTML documents, images, and text. Then the result is used by the search engines. In this paper they have firstly introduced the concepts related to web mining then present an overview of different Web Content Mining tools.Then concluded by presenting a comparative table of these tools based on some pertinent criteria.

In the paper [3] they have described and evaluated methods for learning to forecast forthcoming events of interest from a corpus containing 22 years of news stories. The examples of identifying significant increases in the likelihood of disease outbreaks, deaths and riots in advance of the occurrence of these events in the world. Here the details of methods and studies including the automated extraction and generalization of sequences of events from news corpora and multiple web resources are provided. The predictive power of the approach on real-world events with held from the system is evaluated.

In the paper [4] the author dealed with a preliminary discussion of WEB mining few key computer science contributions in the field of web mining the prominent successful applications and outlines some promising areas of future research. From very beginning the potential of extracting valuable knowledge from the Web has been quite evident. Web mining i.e. the application of data mining techniques to extract knowledge from Web content, structure and usage is the collection of technologies to fulfill this potential. Web mining is the application of data mining techniques to extract knowledge from Web data where at least one of structure or usage data is used in the mining process. Interest in Web mining has grown rapidly in its short existence both in the research and practitioner communities.

In the paper [5] the author has described nature-inspired metaheuristic algorithms especially those based on swarm intelligence have attracted much attention in the last ten years. It describes the fundamentals of firefly algorithm together with a selection of recent publications. The discussion is optimality associated with balancing exploration and exploitation, which is essential for all methodes algorithms. By comparing with intermittent search strategy, the conclusion is that method such as firefly algorithm are better than the optimal intermittent search strategy. Analyzation of algorithms and their implications for higher-dimensional optimization problems is done.

In the paper [6] they have proposed an entirely new dimension towards web page classification using Artificial Neural Networks (ANN).World Wide Web is growing at an uncontrollable rate. Hundreds of thousands of web sites appear every day with the added challenge of keeping the web directories up-to-date.The uncontrolled nature of web presents difficulties for Web page classification.As the number of Internet users is growing, so there is a need for classification of web pages with greater precision in order to present the users with web pages of their desired class. However, web page classification has been accomplished mostly by using textual categorization methods. In this paper they have proposed a novel approach for web page classification that uses the HTML information present in a web page for its classification is done.

In the paper [7] they have introduced that Intelligent Water Drops (IWD) algorithm is adapted for feature selection with Rough Set (RS). Specifically, IWD is used to search for a subset of features based on RS dependency as an evaluation function. The resulting system, called IWDRSFS (Intelligent Water Drops for Rough Set Feature Selection), is evaluated with six benchmark data sets. The performance of IWDRSFS are analysed and compared with those from other methods in the literature. The outcomes indicate that IWDRSFS is able to provide competitive and comparable results. In summary, this study shows that IWD is a useful method for undertaking feature selection problems with RS.

In the paper [8] they propose a genetic algorithm to select best features for Web page classification problem to improve accuracy and run time performance of the classifiers. To determine whether a Web page belongs to a specific class (e.g., a graduate student homepage, a course page, etc.) or not, a classifier needs to have "good" features extracted from the Web pages. As every component in a Web page such as HTML tags and terms can be taken as a feature, dimension of the classification problem becomes too high to be solved by well known classifiers like decision trees, support vector machines, etc. To decrease the feature space, we developed a genetic algorithm that determines the best features for a given set of Web pages. It is found that when features selected by our genetic algorithm are used and a kNN classifier is employed, the accuracy improves up to 96%.

## IV. CONCLUSION

The increase in the amount of information on the Web has caused the need for accurate automated classifiers for Web pages to maintain Web directories and to increase search engine performance. Every tag and every term on each Web page can be considered as a feature there is a need for efficient methods to select best features to reduce feature space of the Web page classification problem. The web classification research with respect to its features and algorithms, we conclude this by summarizing the lessons we have learned from existing research and pointing out future opportunities in web classification. Classification tasks include assigning documents on the basis of subject, function, sentiment, genre, and more. We have studied number of techniques for web page classification but due to the rapid growth of data on internet still there is a need of efficient technique. Which will speed up the web page classification process and give the optimized result.

## V. Acknowledgements

I would like to thank to all the people those who have help me to give the knowledge about these research papers and I thankful to my guide with whose guidance I would have completed my research paper and make it to published, finally I like to thank to all the website and IEEE paper which I have gone through and have refer to create my Review paper successful.